\documentclass[pra,amsfonts,amssymb,amsmath,twocolumn,preprintnumbers]{revtex4}

\usepackage[dvipsnames,dvipsnames]{color}
\usepackage[dvips]{graphics}

\newcommand{\mnras}{MNRAS}
\newcommand{\aap}{A\&A}
\newcommand{\aaps}{A\&A Sup. Ser.}
\newcommand{\jqsrt}{J. Quant. Spectrosc. Radiat. Transf.}
\newcommand{\cpc}{Comp. Phys. Comm.}
\newcommand{\jpd}{J. Phys. D, Appl. Phys.}
\newcommand{\ptrs}{Phil. Tran. Roy. Soc. Lon. A, Math. and Phys. Sci.}
\newcommand{\ps}{Phys. Scrip.}
\newcommand{\jpb}{J. Phys. B, At., Mol., Opt. Phys.}
\newcommand{\ppr}{Plas. Phys. Rep.}
\newcommand{\acp}{AIP~{C}onf.~{P}roc.~No.}

\begin{document}
\preprint{Article is copyright (2007) by the American Physical Society, posted with permission}
\preprint{Reference info is available at http://pra.aps.org/ and http://arxiv.org/ (see left side for document number)}
\setcounter{page}{0}
\thispagestyle{empty}
\begin{center}
\resizebox{\textwidth}{!}{\includegraphics*[12,12][599,750]{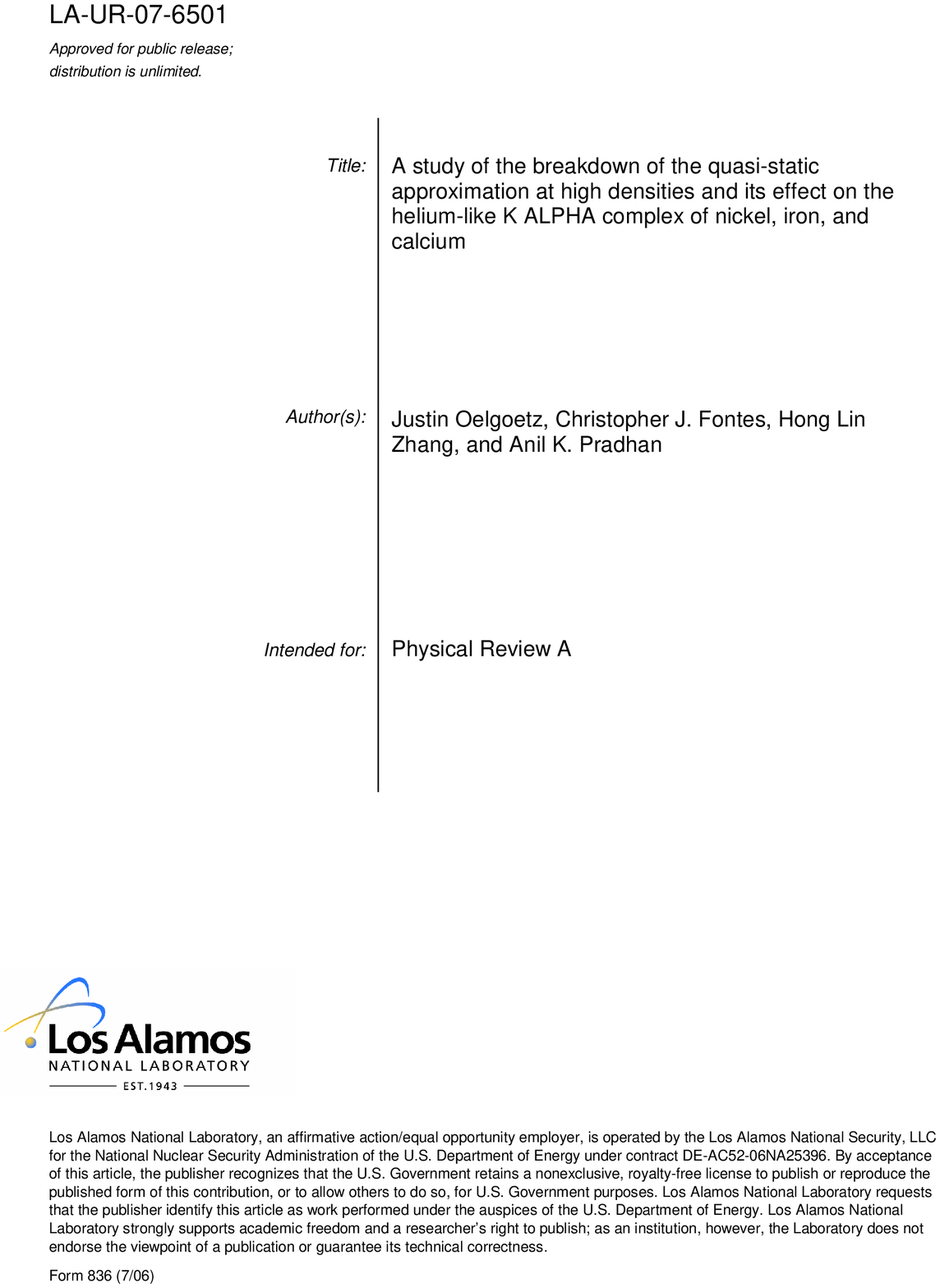}}
\end{center}
\pagebreak

\title{A study of the breakdown of the quasi-static approximation at high densities and its effect on the helium-like K{\boldmath $\alpha$} complex of nickel, iron, and calcium}
\author{Justin Oelgoetz}
\email{oelgoetz@lanl.gov}
\author{Christopher~J.~Fontes}
\author{Hong~Lin Zhang}
\affiliation{Applied Physics Division, Los Alamos National Laboratory, Los Alamos, NM 87545}
\author{Anil~K. Pradhan}
\affiliation{Astronomy Department, The Ohio State University, 140 W. 18th Avenue, Columbus, OH 43210}
\begin{abstract}
Recent work to include R-Matrix data within a larger model comprised mostly of distorted-wave and plane-wave-Born data has resulted in the General Spectral Modeling (GSM) code.  It employs a quasi-static approximation, a standard, low-density methodology that assumes the ionization balance is separable from a determination of the excited-state populations that give rise to the spectra.  GSM further allows for some states to be treated statistically as contributions to effective rates, instead of being included explicitly in the kinetics model.  While these two approximations are known to be valid at low densities, this work investigates using such methods to model high-density, non-LTE emission spectra and determines at what point the approximations break down by comparing to spectra produced by the Los Alamos National Laboratory code ATOMIC which makes no such approximations.  As both approximations are used by other astrophysical and low-density modeling codes, the results should be of broad interest.  He-like K$\alpha$ emission spectra are presented for three elements, Ni, Fe, and Ca, in order to gauge the effect of both the statistical methods and the ground-state-only, quasi-static approximation employed in GSM.  This work confirms that at and above the temperature of maximum abundance of the He-like ionization stage, the range of validity for both approximations is sufficient for modeling the low- and moderate-density regimes one typically finds in astrophysical and magnetically confined fusion plasmas.  However, a breakdown does occur for sufficiently high densities; we obtain quantitative limits that are significantly higher than previous works.  Additionally, this work demonstrates that, while the range of validity for both approximations is sufficient to accurately predict the density-dependent quenching of the z line, the approximations begin to break down at higher densities.  Thus these approximations should be used with greater care when modeling high-density plasmas such as those found in laser-driven inertial confinement fusion and electromagnetic pinch devices.
\end{abstract}

\maketitle

\section{Introduction}

The subject of K$\alpha$ emission lines arising from helium-like ions has been a topic in the literature for the better part of a century \cite{Edlen-Tyren:1939}.  The work that the present effort builds on begins with Gabriel and Jordan's work on the Sun \cite{Gabriel-Jordan:1969-nature,Gabriel-Jordan:1969-helines}.  There are four K$\alpha$ lines arising from the helium-like ionization species, each involving decay from an excited state to the ground state. The `w' line represents a dipole allowed transition arising from the decay of the $1s2p$ $^1P^{o}_1$ state.  Because of mixing, the  $1s2p$ $^3P^{o}_1$ state can also decay via a dipole allowed transition, giving rise to the intercombination line `y'.  The $1s2s$~$^3S_1$ state can decay via a relativistic magnetic-dipole transition, giving rise to the forbidden line, `z'.  Lastly an additional forbidden line, `x', is formed from the $1s2p$~$^3P^{o}_2$ state decaying via a magnetic-quadrupole transition.  Due to the sometimes small energy separation between the $1s2p$ $^3P^{o}_1$ and $1s2p$~$^3P^{o}_2$ states, the x and y lines can not be independently resolved for all elements or plasma conditions.  The $1s2s$ $^1S_0$ state must also be included explicitly in the determination of the excited-state populations as it can decay via a two-photon transition which has an important effect on the model, as does the presence of the $1s2p$~$^3P^{o}_0$ state \cite{Gabriel-Jordan:1969-helines,Mewe-Schrijver:1978stat}.

From these four lines, line ratios have been developed in order to diagnose certain conditions about the corresponding plasma. One of those line ratios, R, is defined as \cite{Gabriel-Jordan:1969-helines}
\begin{equation}
R=\frac{I(\mathrm{z})}{I(\mathrm{x})+I(\mathrm{y})}\;. \label{Rformula}
\end{equation}
The intensities, $I$, of the non-dipole allowed lines (x and z) and the intercombination line (y) vary in a similar manner as a function of temperature.  However, electron collisions can transfer population from the $1s2s$~$^3S_1$ state to the $1s2p$~$^3P^{o}_{0,1,2}$ states, thereby quenching the z line and increasing the intensity of the x and y lines.  Therefore the ratio $R$ can be used as a diagnostic of electron density. 

Building on this earlier work, the study of K$\alpha$ emission from He-like ions was extended to heavier elements by including the effect of dielectronic satellite lines \cite{Gabriel:1972-sats}, and to a broader range of elements and transient conditions in the seminal works of Mewe and Schrijver \cite{Mewe-Schrijver:1978stat,Mewe-Schrijver:1978nonstat}.  Improved data and improved models continued to be brought to bear as they became available \cite{Pradhan-Shull:1981-helines,Pradhan:1982-helines,Pradhan:1985-cascade,Lee-etal:1985,Lee-etal:1986}.  Photoionized media have also been investigated \cite{Porquet-Dubau:2000}.  Lastly, more recent Breit-Pauli R-Matrix \cite{BPRM,Nahar-Pradhan:1994-urr} data were used to refine the quantities of interest even further \cite{Oelgoetz-Pradhan:2001,Oelgoetz-Pradhan:2004}.  Additionally the K$\alpha$ lines are also observed in laboratory plasmas \cite{Edlen-Tyren:1939,Negus-Peacock:1979,Peacock-Burgess:1981,Bitter-etal:1993,Bertschinger-etal:1999,Marchuk-etal:2004,vonHellermann-etal:2005,Safronova-etal:2006}, and these same methods have been successfully used to model and interpret the spectra of magnetically confined fusion plasmas \cite{Bitter-etal:1993,Bertschinger-etal:1999,Marchuk-etal:2004,vonHellermann-etal:2005}.  While the above narrative and citations are only a small portion of the work done in this field since these lines were initially observed, they are the platform upon which the General Spectral Modeling (GSM) code \cite{Oelgoetz-Dissertation}, one of the codes used in the present work, is based.

It should be noted that much of the previous work explicitly assumed that, for the purposes of finding the populations of the He-like excited states, the populations of the adjacent ionization stages are entirely in the ground state.  While this particular type of quasi-static approximation (the ground-state-only, quasi-static approximation) is clearly valid at low densities, it breaks down at higher densities as electron-impact ionization and recombination out of excited states become important mechanisms for populating the excited states.  One could envision extending this treatment by including not only the ground state in the set of states with which the excited states are in equilibrium, but also select metastable states.  The effect of expanding the quasi-static treatment employed in this way is not explored in the present work, as the vast majority of papers on astrophysical He-like K$\alpha$ emission make these same approximations.  Additionally many of the above works treat many of the excited states as cascade corrections (see, for example, \cite{Pradhan:1985-cascade}) to direct rate coefficients.  The effect of such a statistical approximation is explored in the present work.

From a more general perspective, expanding beyond the specific case of He-like ions described above, the trend over time has been to incorporate more accurate atomic data in plasma kinetics and spectral modeling.  Often, these better atomic data are difficult or impossible to calculate for all processes between all levels.  Furthermore, if the atomic data are calculated using a method that accounts for resonance structure, such as the R-Matrix method, it is much more time consuming to calculate rate coefficients from the more complex cross sections than from smooth cross sections.  Reducing the computational requirements makes the quasi-static and statistical approximations appealing.  A separate, but related, consideration is that as the plasma transitions from non-local-thermodynamic-equilibrium (non-LTE) conditions to the high density regime, an accurate modeling approach should produce the appropriate LTE limit.  However, each of the two approximations mentioned above typically preclude a transition to the correct LTE population distribution in the high-density limit, regardless of the quality of the atomic data that is used.  The flexibility of GSM allows for an investigation of the interplay between these various considerations when attempting to model plasmas over a broad range of conditions.

The primary interest of this work is benchmarking and examining the range of validity for the ground-state-only, quasi-static approximation as well as the statistical treatment of excited states in the context of He-like ions, as these approximations are employed by most of the works referenced above.  Many of the connections between excited states are neglected by these two approximations; thus the constraints on the system are not sufficient to produce the correct high-density, local thermodynamic equilibrium (LTE) limit.   Using these approximations results in a significant speed up however, and allows one to include the effects of many more states with the same computing power.  Thus the applicability of these approximations to high-density, non-LTE plasmas such as those found in vacuum sparks \cite{Negus-Peacock:1979}, electromagnetic pinch type devices \cite{Bakshaev-etal:2001,Safronova-etal:2006} and laser produced, inertial confinement fusion plasmas \cite{Peacock-Burgess:1981} is of great interest.  To gauge the effect of these approximations, the spectra produced by GSM are compared to the results from another code, ATOMIC \cite{ATOMIC}.  ATOMIC does not employ either the quasi-static approximation or statistical methods for treating excited states; instead every fine-structure level in the model is treated identically, such that all included processes between all levels are considered explicitly.  It should be noted that the present work is not the first comparison of a code based on the quasi-static approximation to ATOMIC; comparisons between the code ADAS \cite{ADAS} and ATOMIC have been previously considered \cite{Loch-etal.2004}.  However, the primary aim of the present work is to investigate the breakdown of the approximations involved and to provide quantitative limits for the breakdown, while the earlier work considered the effects of differing atomic data sets on plasma kinetics modeling.

The secondary aim of this work is to confirm that using a statistical treatment and the ground-state-only, quasi-static approximation are adequate for modeling emission spectra in the low- and moderate-density regimes encountered in magnetic confinement fusion devices and most astrophysical plasmas.  As the line ratio $R$ is useful for determining the density of the emitting plasma, and many previous calculations of $R$ employ both a statistical treatment of excited states and the ground-state-only, quasi-static approximation, we also confirm that the methods used for such calculations are indeed adequate.  The test cases presented in this work are for Ni, Fe, and Ca as they are directly relevant to astrophysical and laboratory plasmas.

\section{Theory}

Two distinct atomic kinetics codes, ATOMIC \cite{ATOMIC} and GSM \cite{Oelgoetz-Dissertation}, have been used to explore the same plasmas in an effort to determine the range of validity of the ground-state-only, quasi-static approximation and statistical treatment of excited states for the three elements considered in this work.  ATOMIC has evolved within the context of modeling high-density plasmas where assumptions like the quasi-static approximation can be problematic.  As previously mentioned, GSM employs the ground-state-only, quasi-static approximation, and has the ability to treat states statistically.  While GSM enables the use of more accurate atomic parameters, such as R-matrix data, the methods used by GSM are an approximation to the more general approach considered in ATOMIC; thus it is convenient to start with a description of the theory employed by the ATOMIC code and then move on to describing GSM.  

\subsection{ATOMIC}
ATOMIC \cite{ATOMIC} is a general kinetics code which solves the collisional-radiative atomic equations given by
\begin{eqnarray}
\frac{dN_{l,j}}{dt} & = &N_e\left(\sum_{i(i\ne j)}(N_{l,i}q_{i \rightarrow j}(\tilde{\varepsilon})-N_{l,j}q_{j \rightarrow i}(\tilde{\varepsilon}))\right. \nonumber \\
&+&\sum_{i}N_{l+1,i}C_{l+1,i \rightarrow l,j}(\tilde{\varepsilon})-N_{l,j}\sum_{i}C_{l,j \rightarrow l-1,i}(\tilde{\varepsilon})\nonumber \\
 &+&\sum_{i}N_{l-1,i}D^{\mathrm{DC}}_{l-1,i \rightarrow l,j}(\tilde{\varepsilon}) -N_{l,j}\sum_{i}D^{\mathrm{DC}}_{l,j \rightarrow l+1,i}(\tilde{\varepsilon})\nonumber \\ 
&+&\left.\sum_{i}N_{l-1,i}\alpha^{\mathrm{RR}}_{l-1,i \rightarrow l,j}(\tilde{\varepsilon}) -N_{l,j}\sum_{i}\alpha^{\mathrm{RR}}_{l,j \rightarrow l+1,i}(\tilde{\varepsilon})\right)\nonumber \\
&-&N_{l,j}\sum_{i(i<j)}A_{j \rightarrow i}+N_e^2\left(\sum_{i}N_{l-1,i}\beta_{l-1,i \rightarrow l,j}(\tilde{\varepsilon})\right. \nonumber \\
&-&\left.N_{l,j}\sum_{i}\beta_{l,j \rightarrow l+1,i}(\tilde{\varepsilon})\right) +\sum_{i(i>j)}N_{l,i}A_{i \rightarrow j}  \nonumber \\
&+&\sum_{i}N_{l+1,i}R^{\mathrm{AI}}_{l+1,i \rightarrow l,j} - N_{l,j}\sum_{i} R^{\mathrm{AI}}_{l,j \rightarrow i,l-1} \label{atomicformula}\;.
\end{eqnarray}
One can see from Eq. (\ref{atomicformula}) that all states are treated explicitly whether they be bound or autoionizing.  Rate coefficients for electron-impact excitation ($q$), collisional ionization ($C$), dielectronic capture ($D^{\mathrm{DC}}$), radiative recombination ($\alpha^{\mathrm{RR}}$), three-body recombination ($\beta$), radiative decay ($A$), and autoionization ($R^{\mathrm{AI}}$) are included between all possible fine-structure levels in the model.  A variable describing the electron energy distribution is denoted by $\tilde\varepsilon$; as all results presented in this work are for thermal systems, $\tilde\varepsilon$ can be taken to be the electron temperature.  As radiation fields are not considered in this work, processes depending on them have been omitted from the discussion but are in general included by ATOMIC.  This method is guaranteed to go to the correct LTE limit as the electron density increases since the rate coefficients for the inverse processes are calculated assuming detailed-balance with those of the forward processes.

Once the populations ($N_{l,j}$) have been calculated for each level ($j$) of every ionization stage ($l$), emission spectra ($S$) are determined by finding the intensity ($I$) of each line and then applying a Doppler-broadened line profile to it using the equations
\begin{equation}
I(l,j\rightarrow l,k)=N_{l,j}A_{j \rightarrow k}\label{Ieqn}\;,
\end{equation}
\begin{equation}
S(h\nu)=\sum_{l,j,k} I(l,j\rightarrow l,k)h\nu\frac{c\sqrt{m_i}}{2\pi kT_i}\,e^{\frac{m_ic^2(h\nu-\Delta E_{jk})^2}{2\Delta E_{jk}^2kT_i}} \label{Seqn}\;,
\end{equation}
where $m_i$ and $T_i$ are the ion mass and temperature respectively, and $\Delta E_{jk}$ is the energy of the line produced by radiative decay from level $j$ to $k$.

\subsection{GSM}
GSM \cite{Oelgoetz-Dissertation} is based on the quasi-static approximation (see, for example, \cite{Loch-etal.2004}) as implemented by much of the previous work on modeling He-like spectra \cite{Mewe-Schrijver:1978stat,Pradhan-Shull:1981-helines,Pradhan:1982-helines,Oelgoetz-Pradhan:2001,Oelgoetz-Pradhan:2004}.  As such, GSM assumes that the excited states of a particular ionization stage are always in instantaneous equilibrium with the adjacent ionization stages, which are assumed to be entirely in the ground state.  The assumption is valid at low densities since the excitation and radiative decay rates are generally higher than ionization and recombination rates that determine ionization balance.

The result of employing the ground-state-only, quasi-static approximation is that the first step in any calculation is to obtain the ionization fractions by solving the following set of coupled equations
\begin{eqnarray}
\frac{dX_l}{dt}&=&N_e(X_{l+1}\alpha_{l+1 \rightarrow l}(\tilde{\varepsilon})+X_{l-1}C_{l-1 \rightarrow l}(\tilde{\varepsilon})) \nonumber \\ && + N_e^2(X_{l+1}\beta_{l+1 \rightarrow l}(\tilde{\varepsilon}) - X_{l}\beta_{l \rightarrow l-1}(\tilde{\varepsilon})) \nonumber \\ && - X_{l}N_e(\alpha_{l \rightarrow l-1}(\tilde{\varepsilon}) + C_{l \rightarrow l+1}(\tilde{\varepsilon})) \label{ionbaleqn} \;,
\end{eqnarray}
where $X_l$ is the total population in the $l^{th}$ ionization stage, $N_e$ is the electron number density, $C$ is a bulk collisional ionization rate coefficient, $\beta$ is a bulk three-body recombination rate coefficient, and $\alpha$ is a bulk recombination rate coefficient (which includes radiative and dielectronic recombination).  Again, terms involving a radiation field, such as photoionization or stimulated recombination, have been omitted as this work assumed a collision-dominated plasma for which one can neglect the radiation field.  The bulk rate coefficients are typically calculated by summing over the individual pathways involved, or from literature sources that have performed such operations (e.g. \cite{Mazzotta-etal.1998}), and the $X_l$ are obtained by solving the resulting set of coupled equations.  However, in this study the solutions to these equations are taken to be those found by summing over the level populations ($N_{l,j}$) of each ionization stage as determined from an explicit ATOMIC calculation (see Eq. (\ref{atomicformula}) and Fig. \ref{ionbalfig}).  This choice removes the possibility of any discrepancy that might be caused by inconsistencies in the ionization-balance data employed by each code.

GSM offers the additional option to treat a portion of the states in a statistical manner, as conduits or intermediate pathways that are involved in the calculation of the rate coefficients that are used in a reduced system of coupled equations.  The populations of these conduit states are not computed, but all other levels are treated explicitly. The result is a set of effective rate coefficients which, along with the total populations in each ionization stage, are then used to determine the excited-state populations by solving the system of coupled equations given by
\begin{eqnarray}
\frac{dN_{l,j}}{dt} & = &N_e\left( \sum_{i(i\ne j)}(N_{l,i}q^{\mathrm{eff}}_{i \rightarrow j}(\tilde{\varepsilon})-N_{l,j}q^{\mathrm{eff}}_{j \rightarrow i}(\tilde{\varepsilon})) \right.\nonumber \\
&+& X_{l+1}C^{\mathrm{eff}}_{l+1,1 \rightarrow l,j}(\tilde{\varepsilon})-N_{l,j}\sum_{i}C^{\mathrm{eff}}_{l,j \rightarrow l-1,i}(\tilde{\varepsilon}) \nonumber \\ 
&+&\left.  X_{l-1}\alpha^{\mathrm{eff}}_{l-1,1 \rightarrow l,j}(\tilde{\varepsilon}) -N_{l,j}\sum_{i}\alpha^{\mathrm{eff}}_{j,l \rightarrow i,l+1}(\tilde{\varepsilon}) \right)\nonumber \\ 
&+&\sum_{i(i>j)}N_{l,i}A^{\mathrm{eff}}_{i \rightarrow j}+N_e^2\bigg(X_{l-1}\beta^{\mathrm{eff}}_{l-1,1 \rightarrow l,j}(\tilde{\varepsilon}) \nonumber \\
&-&N_{l,j}\sum_{i}\beta^{\mathrm{eff}}_{j,l \rightarrow i,l+1}(\tilde{\varepsilon})\bigg)-N_{l,j}\sum_{i(i<j)}A^{\mathrm{eff}}_{j \rightarrow i} \nonumber \\
&-&N_{l,j}\sum_{i} R^{\mathrm{AI-eff}}_{l,j \rightarrow i,l-1}
 \;,\label{excitedformula} \\
X_l &=& \sum_j N_{l,j}\; ,
\end{eqnarray}
where the terms are defined identically to the case for ATOMIC, except that the superscript `eff' denotes an effective rate coefficient, which includes contributions from levels that do not appear explicitly.   Just as in Eqs. (\ref{atomicformula}) and (\ref{ionbaleqn}), Eq. (\ref{excitedformula}) omits terms that involve the radiation field since it is neglected in this work.

These effective rate coefficients are calculated by summing over the direct and all indirect paths involving states being treated statistically.  The indirect paths involve the use of the {\it collisionless transition matrix} (CTM), $\mathbf{T}_{m \rightarrow j}$, which can be thought of as the probability that an ion in statistical state $m$ will end up in state $j$, via spontaneous processes.  Thus radiative decay and autoionization are the only processes included in calculating the CTM.  It is calculated from these rates using the recursive relation
\begin{eqnarray}
\mathbf{T}_{i\rightarrow j}&=&\sum\limits_{k\not\in Q \atop (E_i>E_k>E_j)}\frac{\Gamma_{i\rightarrow k}}{\sum\limits_lA_{i\rightarrow l}+\sum\limits_m R^{\mathrm{AI}}_{i \rightarrow m}}\mathbf{T}_{k\rightarrow j}\nonumber \\
&+&\frac{\Gamma_{i\rightarrow j}}{\sum\limits_lA_{i\rightarrow l}+\sum\limits_m R^{\mathrm{AI}}_{i\rightarrow m}} \;,
\end{eqnarray}
where $\Gamma_{i\rightarrow k}$ is either a radiative decay rate or an autoionization rate.  The CTM is used in the calculation of all the effective rate coefficients.  For example, effective electron-impact excitation and de-excitation rate coefficients are calculated as
\begin{eqnarray}
q^{\mathrm{eff}}_{j \rightarrow k}(\tilde{\varepsilon}) & = & q^{\mathrm{direct}}_{j \rightarrow k}(\tilde{\varepsilon}) + \sum_{l \atop (E_l>E_j,E_l>E_k)}q^{\mathrm{direct}}_{j \rightarrow l}(\tilde{\varepsilon})\mathbf{T}_{l \rightarrow k} \nonumber \\
&+&  \sum_{i \atop (E_i>E_j,E_i>E_k,E_i>0)}D^{\mathrm{DC}}_{j \rightarrow i}(\tilde{\varepsilon})\mathbf{T}_{i \rightarrow k}\label{eieeq} \;,
\end{eqnarray}
where sums correspond not only to excitation followed by radiative cascade, but also correspond to capture into an autoionizing state followed by autoionization, or some combination of autoionization and radiative decay which produces the final explicit state $k$.  Similar expressions hold for other effective rate coefficients.  Once the effective rate coefficients are used to calculate the level populations via Eq. (\ref{excitedformula}), these populations are used to produce spectra using Eqs.~(\ref{Ieqn}) and (\ref{Seqn}), just as in the ATOMIC calculations.

\section{Computations}
\begin{figure*}
\resizebox{0.89\textwidth}{!}{\includegraphics*{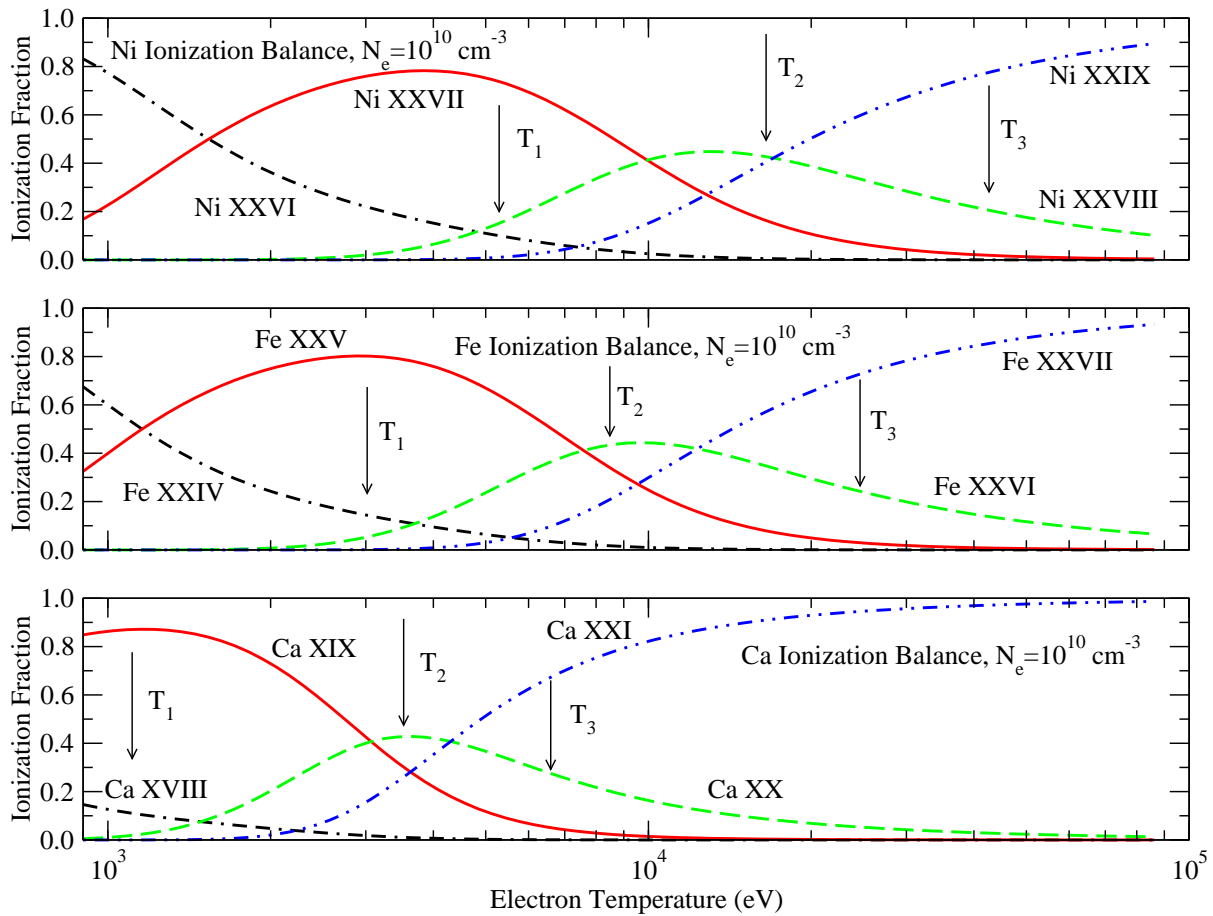}}
\caption{(Color online) Ionization-balance data determined by ATOMIC for the three elements under consideration at an electron density of $N_e=10^{10}$~cm$^{-3}$.  The three temperatures considered for each of the three elements have been indicated.  The first temperature (indicated by the arrow labeled T$_1$) was chosen to be near the temperature of maximum abundance for the He-like ionization stage (the solid red line), the second temperature (indicated by the arrow labeled T$_2$) was chosen to be near the temperature of maximum abundance for the H-like ionization stage (the dashed green line), and the third temperature (indicated by the arrow labeled T$_3$) was chosen such that a significant fraction of the ionization balance consisted of the bare nucleus (the blue dash-dot-dot line).\label{ionbalfig}}
\end{figure*}

In order to simplify the comparisons, GSM was run with the ionization-balance data calculated by ATOMIC, as mentioned earlier in the discussion following Eq. (\ref{ionbaleqn}).  While the ionization balance is a function of both electron temperature and density, the temperatures considered for each element were chosen relative to the low-density limit of the ionization balance (see Fig. \ref{ionbalfig}).  The first temperature ($5.17\times 10^3$~eV ($6.0\times 10^7$~K) for Ni, $3.02\times 10^3$~eV ($3.5 \times 10^7$~K) for Fe, and $1.29\times 10^3$~eV ($1.5 \times 10^7$~K) for Ca) is near the temperature of maximum abundance of the He-like ionization stage.  The second temperature  ($1.72\times 10^4$~eV ($2.0\times10^8$~K) for Ni, $8.62\times 10^3$~eV ($1.0\times10^8$~K) for Fe, and $3.45\times 10^3$~eV ($4.0\times10^7$~K) for Ca) is near the temperature of maximum abundance for the H-like ionization stage.  Lastly the third temperature ($4.31\times 10^4$~eV ($5.0\times10^8$~K) for Ni,  $2.59\times 10^4$~eV ($3.0\times10^8$~K) for Fe, and $6.89\times 10^3$~eV ($8.0\times10^7$~K) for Ca) is significantly above the previous two temperatures and considers the case for which the bare nucleus is dominant.  Furthermore, both ATOMIC and GSM were run with the same set of fundamental atomic data, calculated with the Los Alamos suite of atomic physics codes \cite{Abdallah-etal.1994,Abdallah-etal.2001}.  The CATS code was used to calculate all wave functions, energy levels, dipole allowed radiative decay rates, and plane-wave-Born electron-impact excitation cross sections arising from bound and autoionizing levels of the configurations $nl$, $1snl$, $2lnl'$, $1s^2nl$, $1s2lnl'$, and $1s3lnl'$ with $n\le 10$ and $l\le g$.  The output of CATS was then used by the GIPPER code to calculate autoionization rates and photoionization cross sections in the distorted-wave approximation, as well as collisional ionization cross sections using a scaled hydrogenic approximation which accurately reproduces distorted-wave results for the ionization stages of interest.  The ACE code was used to obtain distorted-wave, electron-impact excitation cross sections for all transitions out of the lowest seven levels of the He-like ionization stage, as well as out of the three levels in the $1s^22l$ complex of the Li-like stage.  Lastly, non-dipole radiative decay rates for the three elements under consideration, i.e. the magnetic-dipole, magnetic-quadrupole, and the two-photon rates mentioned previously, were obtained from Mewe and Schrijver \cite{Mewe-Schrijver:1978stat}.

As described above, two approximations are considered for the GSM calculations: (1) some states can be treated statistically, and (2) the excited states are in instantaneous equilibrium with just the ground states of the appropriate ionization stages.  In order to test the validity of the statistical treatment of certain states, two GSM models were considered.  The first model, called GSM7, treats the lowest seven levels of the He-like ionization stage, along with the ground state of the H-like ionization stage, and the levels arising from the $1s^22l$ and $1s2lnl'$ configurations in the Li-like ionization stage as explicit.  The second model, called GSMF, considers all levels to be explicit.  Thus, this full model includes and accounts for collisions among the excited states within an ionization stage.  It should be noted that due to the second approximation, this full model neglects population mechanisms into these excited states from levels other than the ground state in the adjacent ionization stages.  It does, however, include recombination and ionization from these excited states into all the levels of the adjacent ionization stages as possible mechanisms for population loss via the effective rate coefficients.  Whenever results obtained from the GSMF model differ from those computed with GSM7 model, it is an indication that the excited states can no longer be treated statistically.

Similarly, the validity of the ground-state-only, quasi-static approximation is tested by comparing results from the GSMF model to the output of the ATOMIC code (which is referred to as ATOMIC).  Differences between the GSMF and ATOMIC models indicate that the separation of the ionization-balance and excited-state population calculations is not valid.

\section{Results and Discussion}
\begin{figure*}
\resizebox{\textwidth}{!}{\includegraphics*{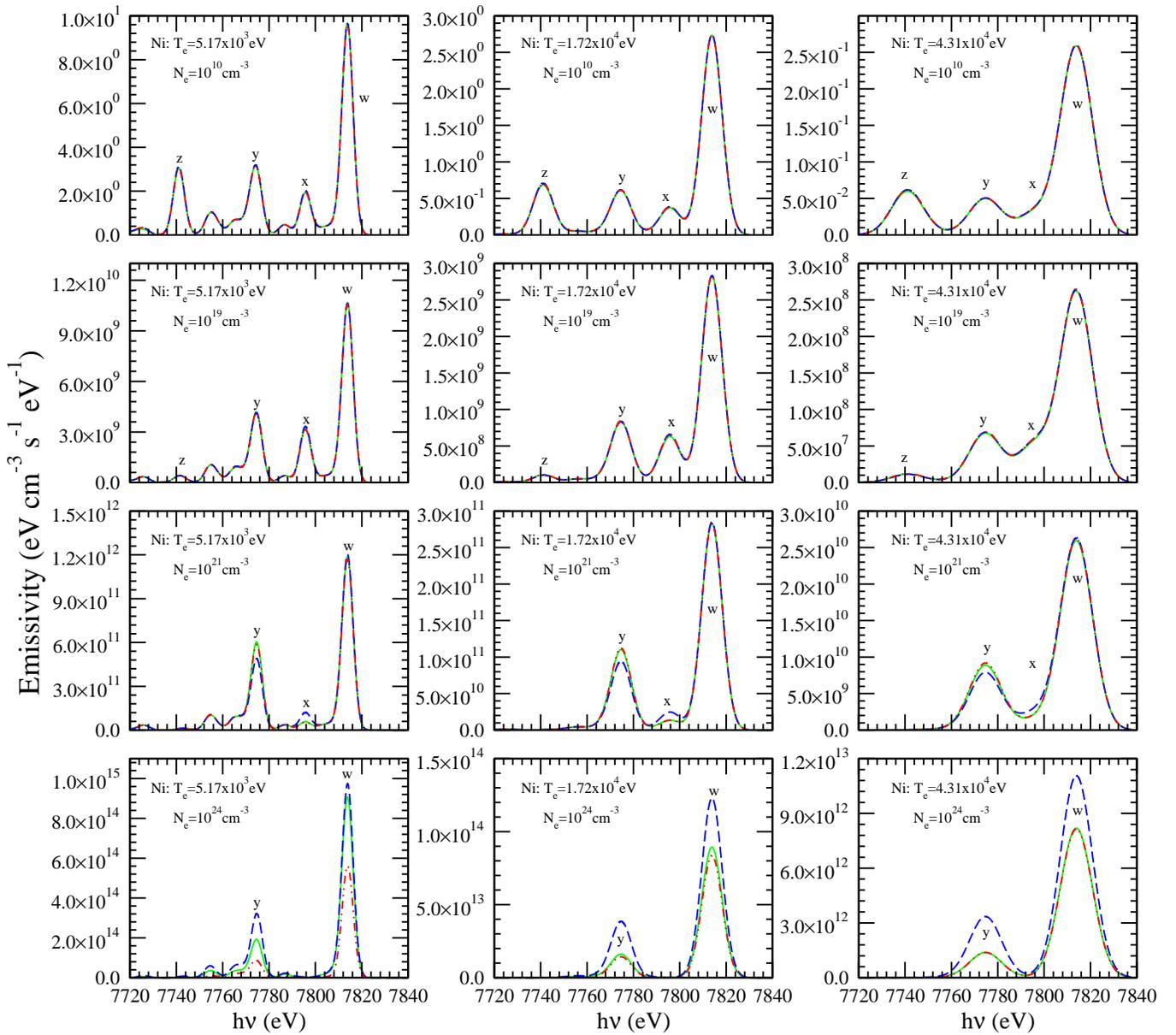}}
\caption{(Color online) Ni emission spectra at T$_e$=$5.17\times10^3$, $1.72\times10^4$, and $4.31\times10^4$~eV ($6.0\times10^7$, $2.0\times10^8$, and $5.0\times10^8$~K) for the three models considered at four different electron densities (top to bottom): $10^{10}$, $10^{19}$, $10^{21}$, and $10^{24}$~cm$^{-3}$.  The solid line (green) was calculated with the ATOMIC model, dashed (blue) with the GSM7 model, and the dot-dot-dashed (red) with the GSMF model.\label{nifig}}
\end{figure*}
\begin{figure*}
\resizebox{\textwidth}{!}{\includegraphics*{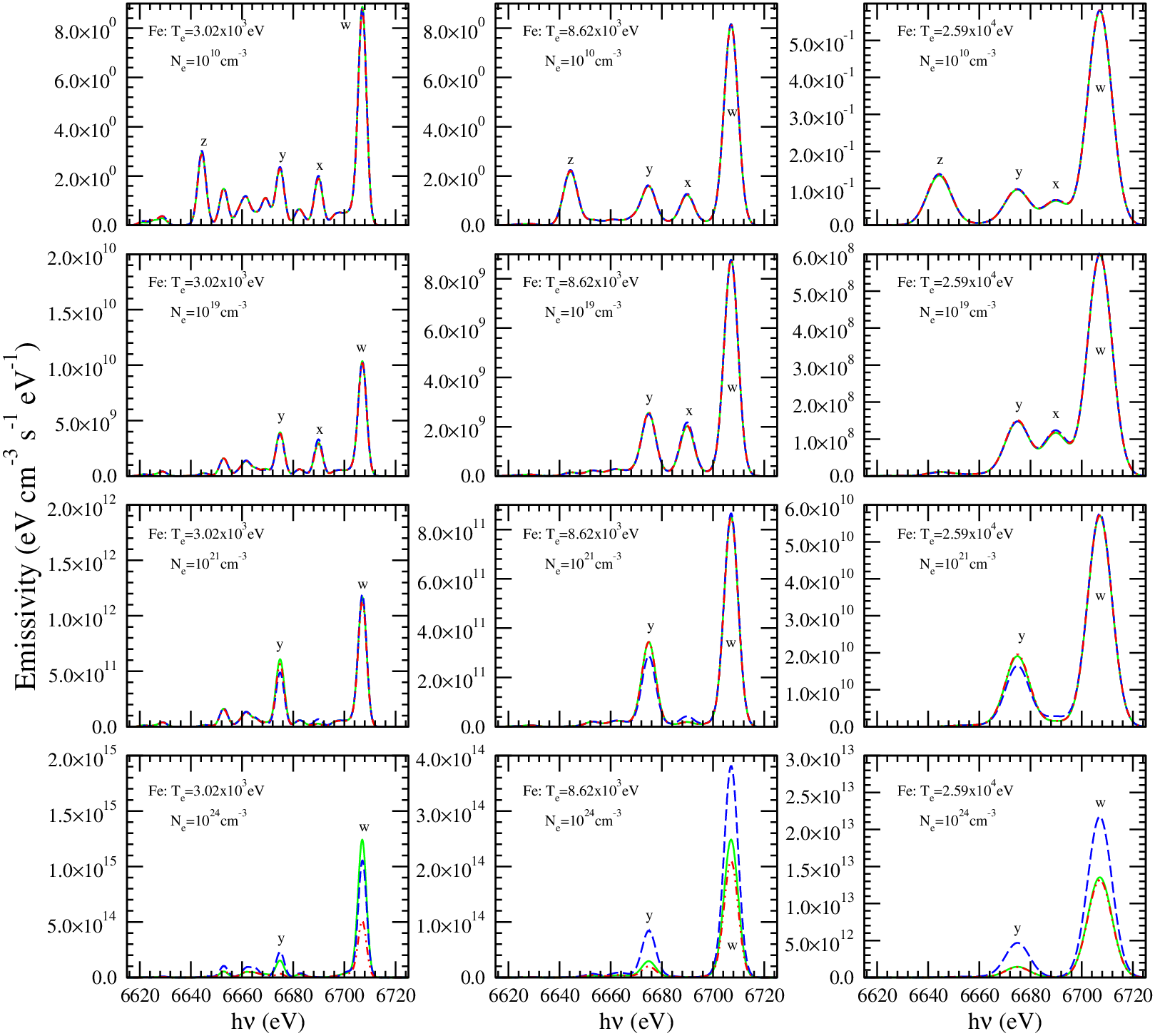}}
\caption{(Color online) Fe emission spectra at T$_e$=$3.02\times10^3$, $8.62\times10^3$, and $2.59\times10^4$~eV ($3.5\times10^7$, $1.0\times10^8$, and $3.0\times10^8$~K) for the three models considered at four different electron densities (top to bottom): $10^{10}$, $10^{19}$, $10^{21}$, and $10^{24}$~cm$^{-3}$.  The solid line (green) was calculated with the ATOMIC model, dashed (blue) with the GSM7 model, and the dot-dot-dashed (red) with the GSMF model.\label{fefig}}
\end{figure*}
\begin{figure*}
\resizebox{\textwidth}{!}{\includegraphics*{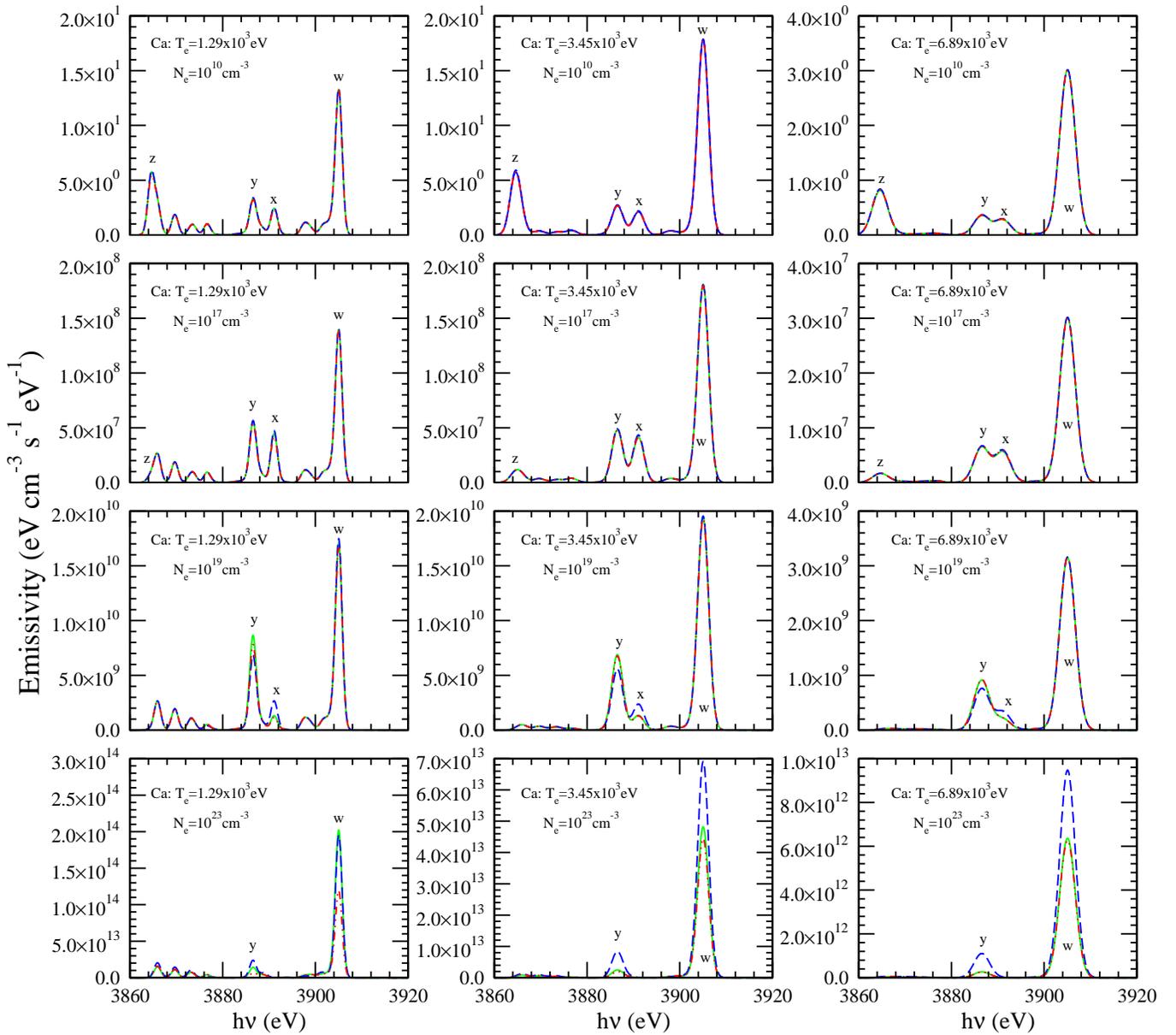}}
\caption{(Color online) Ca emission spectra at T$_e$=$1.29\times10^3$, $3.45\times10^3$, and $6.89\times10^3$~eV ($1.5\times10^7$, $4.0\times10^7$, and $8.0\times10^7$~K) for the three models considered at four different electron densities (top to bottom): $10^{10}$, $10^{17}$, $10^{19}$, and $10^{23}$~cm$^{-3}$.  The solid line (green) was calculated with the ATOMIC model, dashed (blue) with the GSM7 model, and the dot-dot-dashed (red) with the GSMF model.\label{cafig}}
\end{figure*}
\begin{figure*}
\resizebox{0.9\textwidth}{!}{\includegraphics*{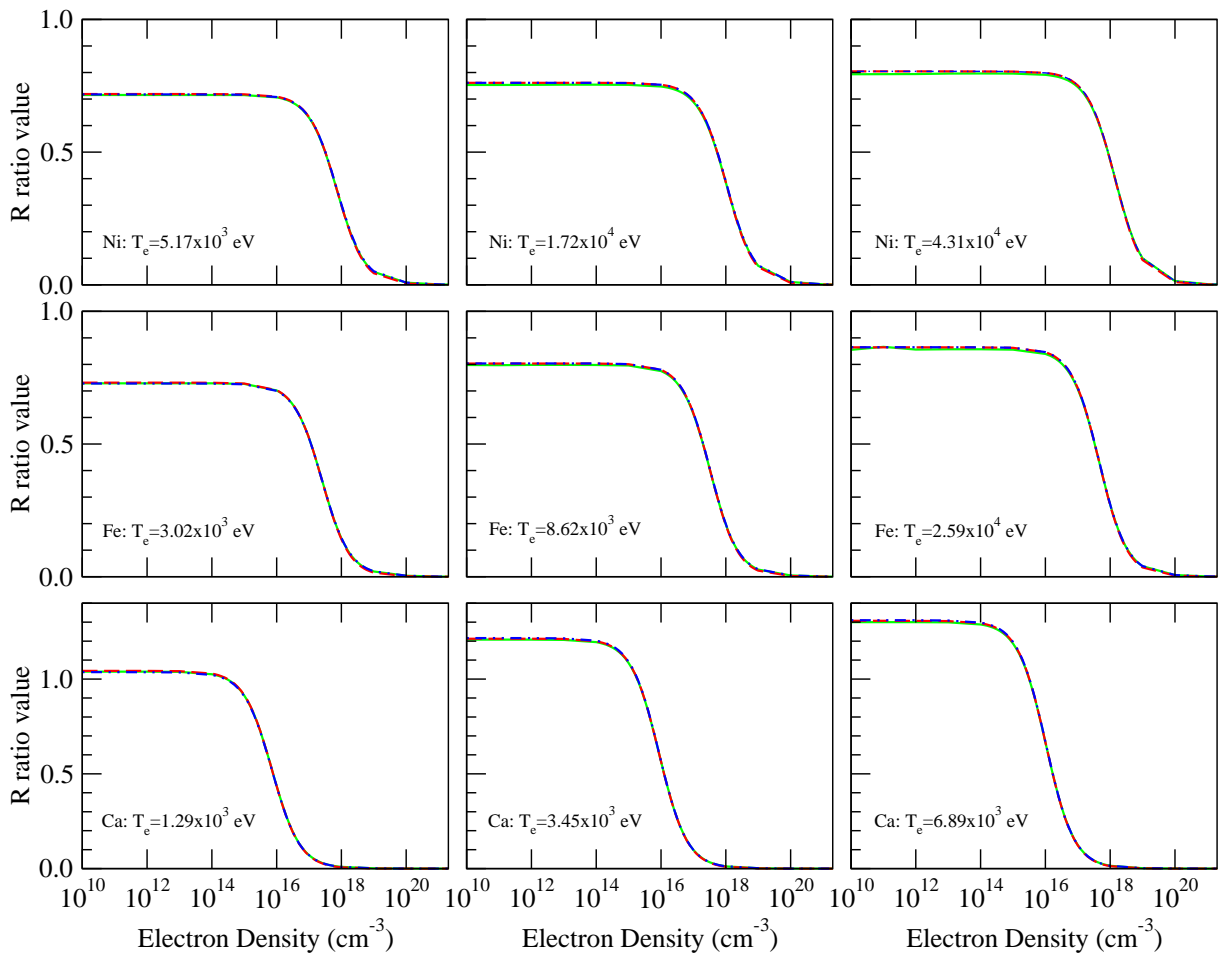}}
\caption{(Color online) A comparison of the K$\alpha$ line ratio $R$ predicted by ATOMIC, and the GSM7 and GSMF models as a function of electron density for the elements and temperatures considered in this work.  The solid line (green) was calculated with the ATOMIC model, dashed (blue) with the GSM7 model, and the dot-dot-dashed (red) with the GSMF model.  Of particular note is the excellent agreement among all three models over the entire range of conditions.\label{rratio}}
\end{figure*}

The first element to be considered is nickel (see Fig. \ref{nifig}).  The top two rows of Fig. \ref{nifig} show emissivities generated from the three models that are in excellent agreement.  The top row of spectra was calculated at a relatively low electron density ($10^{10}$~cm$^{-3}$), typical of astrophysical sources.  The second row of spectra was calculated at a much higher density ($10^{19}$~cm$^{-3}$), but one at which all three models still agree.  This agreement defines a range of validity which persists among all three temperatures.  The unlabeled features are dielectronic satellite lines.  They are formed only at low temperatures, and become weaker at higher densities, most likely due to collisional quenching.  By the third row ($10^{21}$~cm$^{-3}$), the GSM7 model has begun to disagree with the ATOMIC and GSMF models, which is an indication that the statistical treatment of excited states is no longer valid.  In the bottom row ($10^{24}$~cm$^{-3}$) all three models are vastly different at the lowest temperature, an indication that both the ground-state-only, quasi-static approximation and the statistical treatment are no longer valid.  However, the ground-state-only, quasi-static approximation remains valid at the highest temperature.  This trend has a simple explanation: at higher temperatures, ionization out of the Li-like ionization stage plays less of a role in He-like spectral formation.  Even though Li-like metastable states are important fractions of the Li-like ionization stage population, the effect of collisional ionization from them into the He-like ionization stage is not as large of a factor due to the overall decrease in the population of the Li-like ionization stage.  In addition, the populations of the He-like and H-like ionization stages remain dominated by their ground state values.  Thus the ground-state-only, quasi-static approximation remains valid at higher temperatures.  Additionally, the comparatively good agreement between the GSM7 and ATOMIC models at the lowest temperature and highest density is due to a fortuitous cancellation of the effects associated with the approximations employed in the GSM7 model. 
 
The behavior observed for the iron spectra is similar to that of nickel.  While the agreement between the models is again excellent up to an electron density of approximately $10^{19}$~cm$^{-3}$ (top two rows of Fig. \ref{fefig}), the GSM7 results are clearly differing from the other two models at densities of $10^{21}$~cm$^{-3}$ (third row of Fig. \ref{fefig}) for all three temperatures, indicating that the statistical treatment has broken down.  By an electron density of $10^{24}$~cm$^{-3}$ (bottom row of Fig. \ref{fefig}) all three models have diverged for the two lowest temperatures.  Much like the nickel case, no model containing either the statistical treatment or a ground-state-only, quasi-static approximation can be expected to be valid above this density for relatively low temperatures.  However, as in the Ni case, the quasi-static approximation is still valid at the highest temperature because the total population of the Li-like ionization stage is small and the excited-state populations of the He-like and H-like ionization stages are negligible.  The comparatively good agreement between the GSM7 and ATOMIC models for the lowest temperature at the highest density is again fortuitous.  It is worth reiterating that the ground-state-only, quasi-static approximation remains valid for a wider density range at higher temperatures, underscoring the notion that Li-like metastable states are the cause of the breakdown in the ground-state-only, quasi-static approximation at the lower temperatures.  

While the electron densities for which these behaviors are observed in calcium are different, the pattern is once again the same.  Fig.~\ref{cafig} shows the same sort of excellent low-density agreement that was described above for nickel and iron, but by densities of $10^{19}$~cm$^{-3}$ (third row) the GSM7 model has begun to deviate from the others, indicating a breakdown in the statistical methods used.  By densities of $10^{23}$~cm$^{-3}$ (bottom row), all three curves diverge from each other at the lower temperatures, indicating the breakdown of both approximations.  Yet again, the agreement between the ATOMIC and GSMF models at the highest temperature is well understood based on the previous explanation, and the comparatively good agreement between the GSM7 and ATOMIC models for the lowest temperature at this highest density is fortuitous.  The pattern that the ground-state-only, quasi-static approximation is valid over a greater range in density at higher temperatures is also observed in calcium.

With the comparisons from Figs. \ref{nifig}--\ref{cafig} in hand, it is instructive to compare the calculated validity limits with estimates provided in previous works.  While we are not aware of any estimates for the breakdown of the statistical treatment, the ground-state-only, quasi-static approximation has been estimated to be adequate for He-like ions up to a critical electron density of approximately $N_e^{\mathrm{crit}}=6\times10^{12}(Z-2)^{4.3}$~cm$^{-3}$ \cite{Gabriel:1972-sats,Mewe-Schrijver:1978stat}, where $Z$ is the nuclear charge.  Beyond this density the Li-like levels arising from the $1s^22p$ configuration are expected to become important contributors via collisional ionization to the population of the He-like levels that produce the spectral lines of interest.  This expression gives $N_e^{\mathrm{crit}}$ values of approximately $7.2\times 10^{18}$~cm$^{-3}$ for Ni, $5.1\times 10^{18}$~cm$^{-3}$ for Fe, and $1.5 \times 10^{18}$~cm$^{-3}$ for Ca.  The models considered in this work show that the ground-state-only, quasi-static approximation is valid at densities that are one to two orders of magnitude higher than those predicted by this expression for all three elements at the temperatures considered.

Further inspection of the spectra presented in Figs. \ref{nifig}--\ref{cafig} indicates that all three models agree in their predictions of the quenching of the z line as density increases.  In fact, the values for the density-sensitive line ratio $R$ (presented in Fig. \ref{rratio}) predicted from each of the three models are in excellent agreement over a broad range of electron densities.  The statistical methods used by GSM and the ground-state-only, quasi-static approximation do not begin to break down until densities for which the $R$ ratio has a value of nearly zero.  It is worth noting that this density range is adequate to model most astrophysical and magnetic confinement fusion plasmas.

\section{Conclusions}
The ground-state-only, quasi-static approximation and statistical treatments for excited states are often-used methodologies to simplify the spectral modeling of plasmas at low densities.  This work confirms their low-density validity, based on detailed atomic models, and provides numerical values for the electron densities at which these approximations break down for Ni, Fe, and Ca at various temperatures near and above the temperature of maximum abundance for the He-like ionization stage.  At the temperatures considered, the quasi-static approximation using only the ground states of the H-like and Li-like ionization stages appears to be valid for densities that are significantly higher than those predicted by previous works \cite{Gabriel:1972-sats,Mewe-Schrijver:1978stat}.  The ground-state-only, quasi-static approximation is also valid up to densities that are orders of magnitude higher than the density limits calculated for the statistical treatment employed in the GSM7 models in all cases considered.  This work also confirms that the ground-state-only, quasi-static approximation as well as the methods employed to treat excited states as contributions to effective rates via a statistical treatment are valid for predicting the quenching of the z line, and thus the drop in the $R$ ratio, as a function of electron density in the temperature range considered.  However, this study further indicates that these approximations should be used with care at densities where the $R$ ratio would be expected to be close to zero, such as those found in laboratory produced sparks \cite{Negus-Peacock:1979}, high-density electromagnetic pinch devices \cite{Bakshaev-etal:2001,Safronova-etal:2006}, and inertial confinement fusion experiments \cite{Peacock-Burgess:1981}.

Further study is required to determine if the  quasi-static approximation using only the ground states of the H-like and Li-like ionization stages has a smaller range of validity with respect to density than the statistical treatment at temperatures significantly below the temperature of He-like maximum abundance, or for conditions found in transient plasmas.  We do expect the range of validity of the ground-state-only, quasi-static approximation to be narrower in density at temperatures lower than those considered in this work because the population of the Li-like ionization stage should be greater, and thus the effect of the Li-like ionization stage on K$\alpha$ spectral formation via collisional ionization would play a more significant role.

Another avenue of future research concerns the improvement of both the statistical treatment and the quasi-static approximation used in this work without abandoning either approach.  The models that employ statistical methods could be improved by treating more states explicitly and fewer statistically.  Similarly, the quasi-static approximation employed could be improved by allowing for more than just the ground state to be populated in the ionization-balance portion of the calculation.  Investigating both improvements as a function of temperature and density is left to future work.

 
\begin{acknowledgments}
This work was partially conducted under the auspices of the United States Department of Energy at Los Alamos National Laboratory.  Much of the development of GSM was also done at the Ohio Supercomputer Center in Columbus, Ohio.  One of us (AKP) was partially supported by a grant from the NASA Astrophysical Theory Program.
\end{acknowledgments}

\end{document}